\documentclass[preprint,aps,showpacs]{revtex4}

\usepackage{graphicx}
\usepackage{bm}       
\usepackage{mathptmx} 

\begin{document}

\title{CP symmetry and the strong interactions}
\author{Michael Creutz}
\affiliation{
Physics Department, Brookhaven National Laboratory\\
Upton, NY 11973, USA
}

\date{March 2003}                                            

\begin{abstract}
{ I discuss several aspects of CP non-invariance in the strongly
interacting theory of quarks and gluons.  I use a simple effective
Lagrangian technique to map out the region of quark masses where CP
symmetry is spontaneously broken.  I then turn to the possible
explicit CP violation arising from a complex quark mass.  After
summarizing the definition of the renormalized theory as a limit, I
argue that attempts to remove the CP violation by making the lightest
quark mass vanish are not well defined.  I close with some warnings
for lattice simulations.}
\end{abstract}

\maketitle

\section{Introduction}

The SU(3) non-Abelian gauge theory of the strong interactions is quite
remarkable in that, once an arbitrary overall scale is fixed, the only
parameters are the quark masses.  Using only a few pseudo-scalar meson
masses to fix these parameters, the non-Abelian gauge theory
describing quark confining dynamics is unique.  It has been known for
some time \cite{dashen} that, as these parameters are varied from
their physical values, exotic phenomena can occur, including
spontaneous breakdown of CP symmetry.

The possibility of a spontaneous CP violation is most easily
demonstrated in terms of an effective chiral Lagrangian.  In Section
\ref{review} I will review this model for the strong interactions with
three quarks, namely the up, down, and strange quarks.  This lays the
groundwork for the discussion in Section \ref{diagram} of the CP
violating phase.  Section \ref{etaprime} discusses how heavier states,
most particularly the $\eta^\prime$ meson, enter without qualitatively
changing the structure.

Included among the parameters of the strong interactions is a complex
phase which, if present, explicitly violates CP symmetry.  This
parameter appears to be extremely small \cite{dipole} since no such
violation is seen phenomenologically.  A puzzle for grand unification
asks why is CP violation small for the strong interactions but not the
weak \cite{pecceiquinn}.  It is sometimes suggested that a massless up
quark would solve this problem, and I turn to this issue in section
\ref{mup}.  There I argue that the conventional view is incorrect.
The concept of a single massless quark is renormalization scheme
dependent and not physically meaningful.  Indeed, due to
non-perturbative effects, even the sign of a small up-quark mass is
not uniquely defined.  These effects are of higher order in the chiral
expansion.  But as long as the other quark masses are not identically
zero, some ambiguity remains in the definition of the up-quark mass.
This is likely not relevant for most phenomenological issues, but is
unacceptable for solving something fundamental like the strong CP
problem.

Because renormalization is required, the concept of an ``underlying
basic Lagrangian'' does not exist in quantum field theory.  Instead
there are underlying symmetries, and the continuum theory is defined
in terms of those and a few renormalized parameters.  A single
massless quark is not represented by any symmetry, and, because of
confinement, its mass is not appropriate for a renormalized parameter.
To elucidate these points, I review the non-perturbative definition of
the continuum theory and the corresponding ambiguities in the quark
mass.  These issues remain even with the recently discovered chirally
symmetric lattice fermions.  Finally, Section \ref{final} contains
some concluding remarks, including possible impacts of the CP
violating structures for lattice gauge simulations.

\section{The effective model}
\label{review}

A CP violating phase appears naturally in the simplest chiral sigma
model of interacting pseudo-scalar mesons.  In this section I review
the basic model and the standard connections between the quark masses
and the meson masses.  Nothing in this section is new; I am setting
the stage for later discussion.

To be specific, consider the three flavor theory with its approximate
SU(3) symmetry.  Using three flavors simplifies the discussion,
although the CP violating phase can also be demonstrated for the two
flavor theory following the discussion in \cite{mymasspaper}.  I work
with the familiar octet of light pseudo-scalar meson fields
$\pi_\alpha$ with $\alpha=1\ldots8$.  In a standard way (see for
example \cite{km}) I consider an effective field theory defined in
terms of the SU(3) valued group element
\begin{equation}
\label{sigma}
\Sigma=\exp(i\pi_\alpha \lambda_\alpha/f_\pi)\in SU(3).
\end{equation}
Here the $\lambda_\alpha$ are the usual Gell-Mann matrices which
generate the flavor group and $f_\pi$ is a dimensional constant with a
phenomenological value of about 93 MeV.  I follow the normalization
convention that ${\rm Tr} \lambda_\alpha \lambda_\beta =
2\delta_{\alpha\beta}$.  The neutral pion and the eta meson play a
special role in the later discussion; they are the coefficients of the
commuting generators
\begin{equation}
\lambda_3=\pmatrix{1&0&0\cr
0&-1&0\cr 0&0&0\cr}
\end{equation}
and 
\begin{equation}
\lambda_8={1\over \sqrt 3}\pmatrix{1&0&0\cr
0&1&0\cr 0&0&-2\cr},
\end{equation}
respectively.  In the chiral limit of vanishing quark masses, we model
the interactions of the eight massless Goldstone bosons with the
effective Lagrangian density
\begin{equation}
\label{kinetic}
L_0={f_\pi^2\over 4}{\rm Tr}(\partial_\mu \Sigma^\dagger \partial_\mu \Sigma).
\end{equation}
The non-linear constraint of $\Sigma$ onto the group SU(3) makes this
theory non-renormalizable.  It is to be understood only as the
starting point for an expansion of particle interactions in powers of
their masses and momenta.  Expanding Eq.~(\ref{kinetic}) to second
order in the meson fields gives the conventional kinetic terms for our
eight mesons.

This theory is invariant under parity and charge conjugation.  These
operators generate the transformations
\begin{equation}
\matrix{
&P:\ \Sigma\ \rightarrow\ \Sigma^{-1}\cr
&CP:\ \Sigma\ \rightarrow\ \Sigma^{*}\cr
}
\end{equation}
where the operation $*$ refers to complex conjugation.  The eight
meson fields are pseudo-scalars.  The neutral pion and the eta meson
are both even under charge conjugation.

With massless quarks, the underlying quark-gluon theory has a chiral
symmetry under
\begin{equation}
\matrix{
&\psi_L\rightarrow \psi_L g_L\cr
&\psi_R\rightarrow \psi_R g_R.\cr
}
\end{equation}   
Here $(g_L,g_R)$ is in $(SU(3)\times SU(3))$ and $\psi_{L,R}$
represent the chiral components of the quark fields, with flavor
indices understood.  This symmetry is expected to be broken
spontaneously to a vector SU(3) via a vacuum expectation value for
$\overline \psi_L \psi_R$.  This motivates the sigma model through the
identification
\begin{equation}
\langle 0 | \overline \psi_L \psi_R | 0 \rangle \leftrightarrow v \Sigma.
\label{vec}
\end{equation}
The quantity $v$, of dimension mass cubed, characterizes the strength
of the spontaneous breaking of this symmetry.  Thus our effective
field transforms under the chiral symmetry as
\begin{equation}
\Sigma\rightarrow g_L^\dagger \Sigma g_R.
\end{equation} 
Our initial Lagrangian density is the simplest non-trivial expression
invariant under this symmetry.  

The quark masses break the chiral symmetry explicitly.  From the
analogy in Eq.~(\ref{vec}), these are introduced through a 3 by 3 mass
matrix $M$ appearing in a potential term added to the Lagrangian
density
\begin{equation}
L= L_0 - v {\rm Re\ Tr}(\Sigma M).
\end{equation} 
Here $v$ is the same dimensionful factor appearing in Eq.~(\ref{vec}).
The chiral symmetry of our starting theory shows the physical
equivalence of a given mass matrix $M$ with a rotated matrix
$g_R^\dagger M g_L$.  Using this freedom we can put the mass matrix
into a standard form.  I will assume it is diagonal with increasing
eigenvalues
\begin{equation}
M=\pmatrix{ 
m_u & 0 & 0 \cr
0 & m_d & 0 \cr
0 & 0 & m_s \cr
}
\end{equation} 
representing the up, down, and strange quark masses.  Note that this
matrix has both singlet and octet parts under the vector flavor
symmetry
\begin{equation}
M={m_u+m_d+m_s\over 3}
+{m_u-m_d\over 2}\ \lambda_3+{m_u+m_d-2 m_s\over 2\sqrt 3}\ \lambda_8.
\end{equation} 

In general the mass matrix can still be complex.  The chiral symmetry
allows us to move phases between the masses, but the determinant of
$M$ is invariant.  Under charge conjugation the mass term would only
be invariant if $M=M^*$.  If $|M|$ is not real, then its phase is the
famous CP violating parameter usually associated with topological
structure in the gauge fields.  For the moment I take all quark masses
as real.  Since I am looking for spontaneous CP violation, I consider
the case where there is no explicit CP violation.

To lowest order the masses of the pseudo-scalar mesons appear on
expanding the mass term quadratically in the meson fields.  This
generates an effective mass matrix for the eight mesons
\begin{equation}
{\cal M}_{\alpha\beta}\ \propto\ {\rm Re\ Tr}\ \lambda_\alpha\lambda_\beta M.
\end{equation}
The isospin-breaking up-down mass difference plays a crucial role in
the later discussion.  This gives this matrix an off diagonal piece
mixing the $\pi_0$ and the $\eta$
\begin{equation}
{\cal M}_{3,8}\ \propto\ m_u-m_d.
\end{equation}
The eigenvalues of this matrix give the standard mass relations
\begin{equation}
\label{mesons}
\matrix{
 m_{\pi_0}^2 \propto\  {2\over 3} \bigg(m_u+m_d+m_s
-\sqrt{m_u^2+m_d^2+m_s^2-m_um_d-m_um_s-m_dm_s}\bigg)\cr
m_{\pi_+}^2= \  m_{\pi_-}^2\propto m_u+m_d \hfill\cr
m_{K_+}^2= \ m_{K_-}^2\propto m_u+m_s \hfill\cr
m_{K_0}^2= \ m_{\overline K_0}^2\propto m_d+m_s\hfill \cr
m_{\eta}^2 \propto \ {2\over 3} \bigg(m_u+m_d+m_s
+\sqrt{m_u^2+m_d^2+m_s^2-m_um_d-m_um_s-m_dm_s}\bigg).\cr
}
\end{equation}  
Here I label the mesons with their conventional names.

Redundancies in these relations test the validity of the model.  For
example, comparing two expressions for the sum of the three quark
masses
\begin{equation}
{2(m_{\pi_+}^2+m_{K_+}^2+m_{K_0}^2)
\over 3(m_{\eta}^2+m_{\pi_0}^2)} \sim 1.07
\end{equation}  
suggests the symmetry should be good to a few percent.  Further ratios
of meson masses then give estimates for the ratios of the quark
masses \cite{km,weinberg,leutwyler}.  For one such combination, look at
\begin{equation}
{m_u\over m_d}=
{
m_{\pi^+}^2+m_{K_+}^2-m_{K_0}^2\over
m_{\pi^+}^2-m_{K_+}^2+m_{K_0}^2}\sim 0.66\ .
\end{equation}
This particular combination is polluted by electromagnetic effects;
another combination partially cancels such while ignoring small
$m_um_d/m_s$ corrections
\begin{equation} 
{m_u\over
m_d}= { 2m_{\pi^0}^2-m_{\pi^+}^2+m_{K_+}^2-m_{K_0}^2\over
m_{\pi^+}^2-m_{K_+}^2+m_{K_0}^2}\sim 0.55\ .
\label{updown}
\end{equation} 
Later I will comment on a third combination for this ratio.  For the
strange quark, one can take
\begin{equation}
{2 m_s\over m_u+m_d}=
{
m_{K_+}^2+m_{K_0}^2-m_{\pi^+}^2\over
m_{\pi^+}^2
}\sim 26.
\end{equation}

\section{Spontaneous CP violation}
\label{diagram}
So far all this is standard.  Now I vary the quark masses and look for
interesting phenomena.  In particular, I want to find spontaneous
breaking of the CP symmetry.  Normally the $\Sigma$ field fluctuates
around the identity in SU(3).  However, for some values of the quark
masses this ceases to be true.  When the vacuum expectation of
$\Sigma$ deviates from the identity, some of the meson fields acquire
expectation values.  As they are pseudo-scalars, this necessarily
involves a breakdown of parity, as noted by Dashen \cite{dashen}.

To explore this possibility, I concentrate on the lightest meson from
Eq.~(\ref{mesons}), the $\pi_0$.  From Eq.~(\ref{mesons}) we can
calculate the product of the $\pi_0$ and $\eta$ masses
\begin{equation}
\label{prod}
m_{\pi_0}^2m_\eta^2
\ \propto\ m_um_d+m_um_s+m_dm_s.
\end{equation}
Whenever
\begin{equation}
\label{boundary}
m_u={-m_sm_d\over m_s+m_d}
\end{equation}
the $\pi_0$ mass vanishes.  For increasingly negative up-quark masses,
our simple expansion around vanishing pseudo-scalar meson fields
fails.  The vacuum is no longer approximated by fluctuations of
$\Sigma$ around the unit matrix; instead it fluctuates about an SU(3)
matrix of form
\begin{equation}
\Sigma=\pmatrix{
e^{i\phi_1}&0&0\cr
0&e^{i\phi_2}&0\cr
0&0&e^{-i\phi_1-i\phi_2}\cr
}
\end{equation}
where the phases satisfy
\begin{equation}
m_u \sin(\phi_1)=m_d\sin(\phi_2)=-m_s\sin(\phi_1+\phi_2).
\end{equation}
There are two minimum action solutions, differing by flipping the
signs of these angles.  The transition is a continuous one, with
$\Sigma$ going smoothly to the identity as the boundary given by
Eq.~(\ref{boundary}) is approached.

\begin{figure*}
\centering
\includegraphics[width=3in]{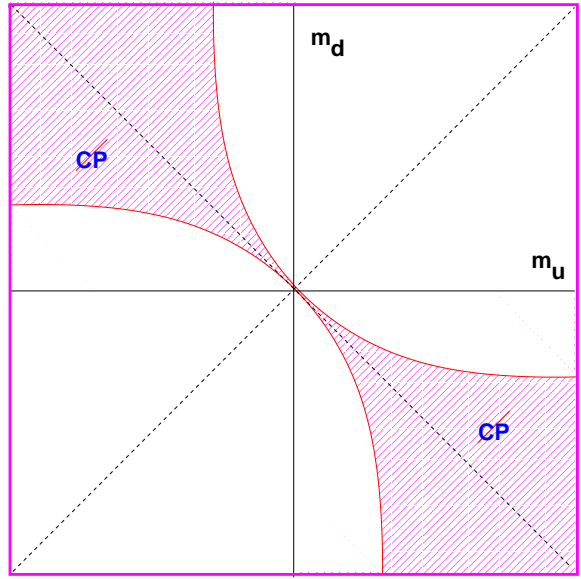}
\caption{\label{fig:phasediagram} The phase diagram of quark-gluon
dynamics as a function of the two lightest quark masses.  The shaded
region exhibits spontaneous CP breaking.  The diagonal lines with
$m_u=\pm m_d$ trace where we have three degenerate pions due to
isospin symmetry.  The neutral pion mass vanishes on the boundary of
the CP violating phase.}
\end{figure*}

In the new vacuum the neutral pseudo-scalar meson fields acquire
expectation values.  As the neutral pion is CP odd, we spontaneously
break this symmetry.  This will have various experimental
consequences, for example eta decay into two pions becomes allowed
since a virtual third pion can be absorbed by the vacuum.
Fig.~(\ref{fig:phasediagram}) sketches the inferred phase diagram as a
function of the up and down quark masses.  Chiral rotations insure a
symmetry under the flipping of the signs of both quark masses.

At first sight the appearance of the CP violating phase at negative up
quark mass may seem surprising.  Naively in perturbation theory the
sign of a fermion mass can be rotated away by a redefinition
$\psi\rightarrow \gamma_5 \psi$.  However this rotation is anomalous,
making the sign of the quark mass observable.  A more general complex
phase in the mass would also have physical consequences, i.e.
explicit CP violation.  With real quark masses the underlying
Lagrangian is CP invariant, but the above discussion shows that there
exists a large region where the ground state spontaneously breaks this
symmetry.

Vafa and Witten \cite{vafawitten} argued on rather general conditions
that CP could not be spontaneously broken in the strong interactions.
However their argument makes positivity assumptions on the path
integral measure.  When a quark mass is negative, the fermion
determinant need not be positive for all gauge configurations; in this
case their assumptions fail.

The possible existence of this phase was anticipated on the lattice
some time ago by Aoki \cite{aoki}.  For the one flavor case he found
this parity breaking phase with Wilson lattice gauge fermions.  He
went on to discuss also two flavors, finding both flavor and parity
symmetry breaking.  The latter case is now regarded as a lattice
artifact of Wilson fermions.  For a review of these issues see
\cite{myreview}.

In conventional discussions of CP non-invariance in the strong
interactions \cite{witten} appears a phase $e^{i\theta}$ appearing on
tunneling between topologically distinct gauge field configurations.
The famous U(1) anomaly formally allows us to move this phase into the
determinant of the quark mass matrix.  After rotating all phases into
the up-quark mass, we see that our spontaneous breaking of CP is
occurring at an angle $\theta=\pi$.  

A crucial observation is that when the down quark mass is positive,
the CP violating phase does not appear for up-quark masses greater
than a non-zero minimum value.  There exists a finite gap with
$\theta=\pi$ without this symmetry breaking.  The chiral model
predicts a smooth behavior as the up-quark mass passes through zero.
This is the main motivation for the discussion in section \ref{mup} on
the meaning of a vanishing up-quark mass.  Indeed, from the effective
Lagrangian point of view, the real and imaginary parts of the quark
mass are independent parameters.  The absence of experimental evidence
for strong CP violation suggests that the imaginary part of the quark
mass matrix vanishes, but says nothing about the real part.

An interesting special case occurs when the up and down quarks have
the same magnitude but opposite sign for their masses, i.e.
$m_u=-m_d$.  In this situation it is illuminating to rotate the minus
sign into the phase of the strange quark.  Then the up and down quark
are degenerate, and we have restored an exact vector $SU(2)$ flavor
symmetry.  The excitation spectrum will show three degenerate pions,
but they will not be massless due to what might be thought of a vacuum
condensate of eta particles.

\section{Including the $\eta^\prime$}
\label{etaprime}

The above discussion was entirely in terms of the pseudo-scalar mesons
that become Goldstone bosons in the chiral limit.  One might wonder
how higher states can influence this phase structure.  Of particular
concern is the $\eta^\prime$ meson associated with the anomalous
$U(1)$ symmetry present in the classical quark-gluon Lagrangian.
Non-perturbative processes, including topologically non-trivial gauge
field configurations, are well known to generate a mass for this
particle.  I will now argue that, while this state can shift masses
due to mixing with the lighter mesons, it does not make a qualitative
difference in the existence of a phase with spontaneous CP violation.

The easiest way to introduce the $\eta^\prime$ into the effective
theory is to promote the group element $\Sigma$ to an element of
$U(3)$ via an overall phase factor.  Thus I generalize
Eq.~(\ref{sigma}) to
\begin{equation}
\Sigma=\exp\left(i\pi_\alpha \lambda_\alpha/f_\pi
+i\sqrt{ 2\over 3}\eta^\prime/f_\pi\right)
\in U(3).
\end{equation}
The factor $\sqrt{2/3}$ gives the $\eta^\prime$ field the same
normalization as the $\pi$ fields.  Our starting kinetic Lagrangian in
Eq.~(\ref{kinetic}) would have this particle also be massless.  One
way to fix this deficiency is to mimic the anomaly with a term
proportional to the determinant of $\Sigma$
\begin{equation}
L_0={f_\pi^2\over 4}{\rm Tr}(\partial_\mu \Sigma^\dagger \partial_\mu \Sigma)
-C|\Sigma|.
\end{equation}
The parameter $C$ parameterizes the strength of the anomaly in the
$U(1)$ factor.

Now if we include the mass term exactly as before, additional mixing occurs
between the $\eta^\prime$, the $\pi_0$, and the $\eta$.  The
corresponding mixing matrix takes the form
\begin{equation}
\pmatrix{
m_u+m_d & {m_u-m_d \over \sqrt 3} & \sqrt{2\over 3}(m_u-m_d) \cr
{m_u-m_d \over \sqrt 3} & {m_u+m_d+4m_s \over 3} 
          &  {\sqrt 2(m_u+m_d-2m_s)\over 3}\cr
 \sqrt{2\over 3}( m_u-m_d)&  {\sqrt2 (m_u+m_d-2m_s)\over 3} & {2\over 3}m_a \cr
}
\end{equation}
where $m_a$ characterizes the contribution of the non-perturbative
physics to the $\eta^\prime$ mass.  This should have a value of order
the strong interaction scale; in particular, it should be large
compared to at least the up and down quark masses.  The two by two
matrix in the upper left of this expression is exactly what is
diagonalized to find the neutral pion and eta masses in
Eq.~(\ref{mesons}).

The boundary of the CP violating phase occurs where the determinant of
this matrix vanishes.  This modifies Eq.~(\ref{prod}) to
\begin{equation}
m_{\pi_0}^2m_\eta^2 m_{\eta^\prime}^2
\ \propto\ 
m_a (m_um_d+m_um_s+m_dm_s)-
m_u(m_d-m_s)^2-m_d(m_u-m_s)^2-m_s(m_u-m_d)^2.
\end{equation}
The boundary shifts slightly from the earlier result, but still passes
through the origin, leaving Fig.~(\ref{fig:phasediagram})
qualitatively unchanged.

\section{Can the up quark be massless?}
\label{mup}
A oft proposed solution to the strong CP problem \cite{banks,
leutwyler0, cohen, fleming} asks whether $m_u=0$.  From the effective
Lagrangian point of view, this appears to be an artificial setting of
two parameters to zero, the real and imaginary parts of the quark
mass.  The earlier discussion shows that nothing special is expected
to happen as the real part of the quark mass goes through zero.  It is
only the imaginary part that should vanish for CP to be a good
symmetry, at least when the up-quark mass is larger than the value
giving spontaneous breaking.  But speculations on a vanishing up-quark
mass continue, so it is interesting to ask if this can be given
physical meaning.  In this section I investigate precisely what is
meant by a quark mass, and what $m_u=0$ would mean.  I will conclude
that the question of whether $m_u$ could vanish is ill posed.  This is
not relevant for most phenomenological purposes, but is unacceptable
for solving something fundamental, like the strong CP problem.
Ref. \cite{banks} raises some of these issues, pursuing $m_u=0$ anyway
as an accidental symmetry.

The conventional description of this phenomenon changes variables from
the complex quark mass to polar coordinates involving the magnitude of
the up-quark mass and its phase.  I argue below that non-perturbative
effects give a scheme-dependent additive shift in the real part of the
up-quark mass.  Such a shift exposes the singular nature of this
selection of coordinates.  To emphasize the point, imagine changing
variables to the magnitude and phase of $(m_u-4\ {\rm MeV})$.  If the
up-quark mass is 4 MeV, then in such coordinates the theory no longer
depends on the phase.  From this point of view $m_u=4\ {\rm MeV}$ is
an equally good solution to the strong CP issue as is $m_u=0$.

While phenomenology, i.e. Eq.~(\ref{updown}), seems to suggest that
the up quark is not massless, there remains considerable freedom in
extracting that ratio from the masses of the pseudo-scalar mesons.
From Eq.~(\ref{mesons}), the sum of the squares of the $\eta$ and
$\pi_0$ masses should be proportional to the sum of the three quark
masses.  Subtracting off the neutral kaon mass should leave just the
up quark.  Thus motivated, look at
\begin{equation}
{m_u\over m_d}=
{
 3(m_{\eta}^2+m_{\pi_0}^2)/2-2m_{K_0}^2
\over
 m_{\pi^+}^2-m_{K_+}^2+m_{K_0}^2
}
\sim-0.8\ .
\end{equation}
Thus even the sign of the up-quark mass is ambiguous.  This example is
perhaps a bit extreme since it ignores the influence of mixing with
the eta prime in the numerator.  But it shows that there does exist a
substantial phenomenological uncertainty in the quark masses.  More
formal attempts to extend the naive quark mass ratio estimates to
higher orders in the chiral expansion have shown fundamental
ambiguities in the definition of the quark masses \cite{km}.

If two quark masses were to vanish simultaneously, then we would have
exactly massless pions, Goldstone bosons for the resulting flavored
chiral symmetry.  In this case the concept of zero quark mass has
definite physical consequences.  But here I concentrate on whether the
concept of only a single massless quark has any meaning.  While I
could carry along the baggage of the heavier quarks, let me simplify
the discussion and consider the theory reduced to a single flavor of
quark.

As stated in the introduction, because renormalization is required,
the continuum concept of an ``underlying basic Lagrangian'' does not
exist.  The continuum theory is specified in terms of basic symmetries
and a few renormalized parameters.  Because of anomalies, a single
massless quark does not correspond to any symmetry.  In practice the
definition of a field theory relies on a limiting process from a
cutoff version.  As the lattice is the only well understood
non-perturbative cutoff, it provides the most natural framework for
such a definition.  But any regulator must accommodate the known
chiral anomalies, and thus there must be chiral symmetry breaking
terms in the cutoff theory.  These chiral breaking effects come in
many guises.  With a Pauli-Villars scheme, there is a heavy regulator
field.  With dimensional regularization the anomaly is hidden in the
fermionic measure.  For Wilson lattice gauge theory there is the
famous Wilson term.  With domain wall fermions there is a residual
mass from a finite fifth dimension.  With overlap fermions things are
hidden in a combination of the measure and a certain non-uniqueness of
the operator.  I will return to this last case shortly.

The lattice regulator involves introducing a dimensionful parameter,
the lattice spacing $a$.  This feature is not special to the lattice.
The scale anomaly is responsible for masses of hadrons such as the
proton and glueballs, even in the massless quark limit.  For such
physics, any complete regulator must introduce a scale.

The renormalization process tunes all relevant bare parameters as a
function of the cutoff to fix a set of renormalized quantities.  In
the case of the strong interactions, the bare gauge coupling is driven
to zero by asymptotic freedom.  Its cutoff dependence is absorbed into
an overall scale in a well known way via the phenomenon of dimensional
transmutation \cite{cw}.  The only other parameters of the strong
interactions are the quark masses.  For these one inputs a few
particle masses to finally determine the continuum theory uniquely.
For the three flavor theory the most natural observables to fix these
parameters are the masses of the pseudo-scalar mesons.

In the one flavor theory there are no Goldstone bosons, but massive
mesons and baryons should exist.  I need some physical parameter with
which to carry out the renormalization of the quark mass.  For this
purpose I choose the ratio of the lightest boson mass to the lightest
baryon mass.  As both are expected to be stable, this precludes any
ambiguity from particle widths.  Calling the lightest boson the $\eta$
and the baryon $p$, I define
\begin{equation}
r={m_\eta \over m_p}.
\end{equation} 
I expect to be able to adjust this parameter via the quark mass, which
should be tuned to give the desired value.  It should be possible to
give this ratio any value throughout the range from $r=0$ at the
boundary of the above CP violating phase to $2/3$ in the heavy quark
limit.

\begin{figure*}
\centering
\includegraphics[width=4in]{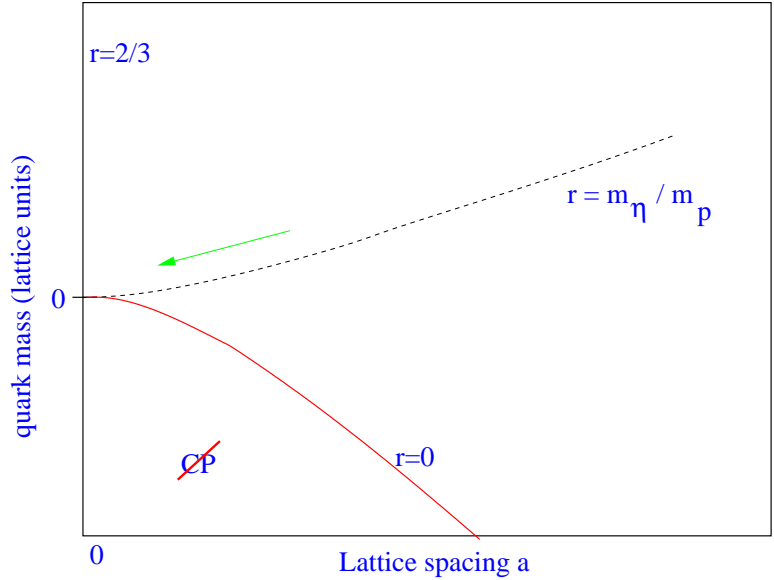}
\caption{
\label{limit}
Defining the continuum limit.  For one-flavor strong interactions I
consider the ratio of the lightest boson to lightest baryon masses as
my renormalized parameter.  With the cutoff in place, we flow towards
the origin along a curve of constant renormalized quantity.  Below the
contour where this ratio vanishes lies the region of spontaneous CP
violation.}
\end{figure*}

With a cutoff in place, I can in principle determine this ratio given
any values for the bare quark mass and bare coupling.  For pedagogy,
let me assume a lattice regulator and trade these parameters for the
lattice spacing $a$ and the quark mass in lattice units, $m_q a$.
Both of these quantities go to zero in the continuum limit.  The
renormalization prescription is to select a desired value of $r$ and
follow the contour with this value in the $(a,m_q)$ plane towards the
origin.  This process is sketched in Fig.~(\ref{limit}).  Perturbative
divergences in the bare quark mass appear in the fact that these
contours approach the origin with zero slope.

When the cutoff is in place, we expect prescription dependent
artifacts.  In particular, the precise locations of the constant $r$
contours will depend on details of the formulation.  Holding the bare
quark mass at zero will cross a variety of $r$ contours, with none
obviously favored as the origin is approached.  Different cutoff
schemes will give different continuum limits for $m_u=0$.
Alternatively, $m_u=0$ in one scheme could give the same continuum
limit as some other scheme with a non-vanishing $m_u$.  A
non-perturbative additive shift in the up-quark mass distinguishes the
two schemes.  Asking that the up-quark mass vanishes is unphysical.

With two or more degenerate flavors there will be one special contour
where the lightest meson does represent a Goldstone boson.  With the
Wilson fermion formulation, the quark mass axis is represented by the
hopping parameter.  As this particular cutoff explicitly breaks chiral
symmetry, the critical hopping parameter, where the meson mass
vanishes, is renormalized away from its value in the continuum limit.

Recently there has been considerable progress with lattice fermion
formulations that preserve a remnant of exact chiral symmetry
\cite{myreview}.  With such, the two flavor theory will have the $r=0$
contour naturally preserved as the $m_q=0$ axis.  The crucial point is
that this is not true for the one flavor theory.  Dynamics generates a
mass for the pseudo-scalar meson; thus, the $m_q=0$ axis will cut
through various finite values of $r$.  An interesting question is
whether, as we take the lattice spacing to zero along this axis, some
physical value of $r$ will be picked out as special and corresponding
to vanishing quark mass.  That this is unlikely follows from the
non-uniqueness of these chiral lattice operators.  For example, the
overlap operator \cite{overlap} is constructed by a projection process
from the conventional Wilson lattice operator.  The latter has a mass
parameter which is to be chosen in a particular domain.  On changing
this parameter, the massless Dirac operator still satisfies the
Ginsparg-Wilson relation \cite{gw}, but this condition does not
guarantee that the contours of constant $r$ in Fig.~(\ref{limit}) will
not shift.  Thus the horizontal axis is not expected to select one
contour as special.  Again, holding $m_u=0$ is not expected to give a
unique continuum theory.

To see that this non-universality is indeed expected, consider that
the dynamics of the one flavor case generates a mass gap in the $\eta$
channel.  This means that the eigenvalues of the Dirac operator
important to low energy physics are are not near the origin, but
dynamically driven a finite distance away.  Changing the projection
procedure to generate the overlap operator will modify the size of
this gap, changing the $\eta$ mass and the ratio $r$.

One might attempt to define $m_u=0$ as the point where the topological
susceptibility vanishes.  For this purpose, the overlap operator
provides a natural definition of the gauge field topology.  With this
prescription a vanishing topological susceptibility is synonymous with
vanishing quark mass for that particular overlap operator.  As the
result depends on the operator chosen, with one flavor the
susceptibility is no more physical than the quark mass.  As a rather
abstract concept, the topological susceptibility is not directly
measurable in scattering processes for physical particles.

This non-perturbative ambiguity in the quark mass carries over to the
explicitly CP violating case where the mass is complex with phase
$\theta$.  The above discussion shows that even the sign of the up
quark mass is ambiguous, indicating that different schemes can have an
ambiguity of $\theta$ between 0 and $\pi$, a particularly severe
example.  For other values of $\theta$, to fix the continuum theory
uniquely we need to introduce another renormalized quantity.  For
example, this could be a three meson coupling or the electric dipole
moment of a baryon.  Beyond the lowest order in the chiral expansion,
the precise dependence of the renormalized parameter on $\theta$ is
scheme-dependent.

\begin{figure*}
\centering
\includegraphics[width=2.5in]{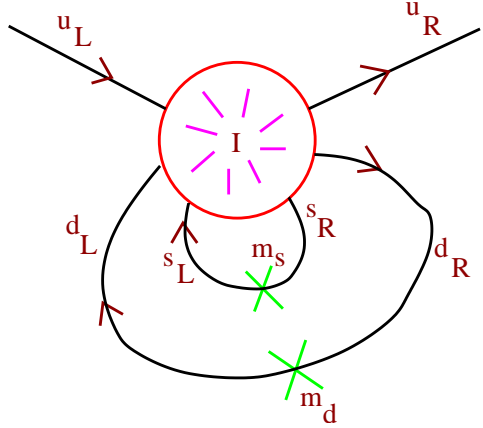}
\caption{
\label{instanton}
Non-perturbative classical gauge configurations can generate an
effective mass term for the up quark.  The magnitude of this mass is
proportional to the product of the heavier quark masses.  The precise
value, however, is scheme
and scale dependent.}
\end{figure*} 

While the use of the lattice provides a framework for precise
discussion, the source of the additive shift in the up-quark mass
follows qualitatively from non-perturbative classical gauge
configurations, i.e. ``pseudoparticles'' or
``instantons'' \cite{mcarthurgeorgi}.  As shown some time ago by 't
Hooft \cite{thooft}, these configurations generate an effective
multi-fermion vertex where all flavors of quark flip their spin.  If
we take this vertex and tie together the massive quark lines with mass
terms, then the resulting process generates an effective mass term for
the light quark.  The strength of this term is proportional to the
product of the masses of the more massive quarks.  This process is
illustrated in Fig.~(\ref{instanton}).  As the strength of the
effective interaction is scheme and scale dependent, so is the
resulting light quark mass.

\section{Final remarks}
\label{final}

While I have been exploring rather unphysical regions in parameter
space, these observations do raise some issues for practical lattice
calculations of hadronic physics.  Current simulations are done at
relatively heavy values for the quark masses.  This is because the
known fermion algorithms tend to converge rather slowly at light quark
masses.  Extrapolations by several tens of MeV are needed to reach
physical quark masses, and these extrapolations tend to be made in the
context of chiral perturbation theory.  The presence of a CP violating
phase quite near the physical values for the quark masses indicates a
strong variation in the vacuum state with a rather small change in the
up-quark mass; indeed, less than a 10 MeV change in the traditionally
determined up-quark mass can drastically change the low energy
spectrum.  Most simulations consider degenerate quarks, and chiral
extrapolations so far have been quite successful.  But some
quantities, namely certain baryonic properties \cite{thomas}, do seem
to require rather strong variations as the chiral limit is approached.
These effects and the strong dependence on the up-quark mass may be
related.

Another issue is the validity of current simulation algorithms with
non-degenerate quarks.  With an even number of degenerate flavors the
fermion determinant is positive and can contribute to a measure for
Monte Carlo simulations.  With light non-degenerate quarks the
positivity of this determinant is not guaranteed.  Indeed, the CP
violation can occur only when the fermions contribute large phases to
the path integral.  Current algorithms for dealing with non-degenerate
quarks \cite{milc} take a root of the determinant with multiple
flavors.  In this process any possible phases are ignored.  Such an
algorithm is incapable of seeing any of the CP violating phenomena
discussed here.  This point may not be too serious in practice since
the up and down quarks are nearly degenerate and the strange quark is
fairly heavy.  But these issues should serve as a warning that things
might not work as well as we want.

\section*{Acknowledgements}
I have benefited from lively discussions with many colleagues, in
particular F. Berruto, T. Blum, H. Neuberger and G. Senjanovic.  This
manuscript has been authored under contract number DE-AC02-98CH10886
with the U.S.~Department of Energy.  Accordingly, the U.S. Government
retains a non-exclusive, royalty-free license to publish or reproduce
the published form of this contribution, or allow others to do so, for
U.S.~Government purposes.

\end{document}